\documentclass[aps,prd,amsfonts,groupedaddress,showpacs,showkeys]{revtex4}
\usepackage{epsfig}
\newcommand{\beaa}{\begin{eqnarray*}} 
\newcommand{\enaa}{\end{eqnarray*}}
\newcommand{\bea}{\begin{eqnarray}}
\newcommand{\ena}{\end{eqnarray}}
\newcommand{\eq}{\begin{eqnarray}} 
\newcommand{\en}{\end{eqnarray}}
\newcommand{\ra}{\rangle}
\newcommand{\la}{\langle}

\begin{document}

\title{Strong decays of radially excited mesons in a chiral approach} 

\author{Thomas Gutsche, Valery E. Lyubovitskij\footnote{On leave of 
absence from the Department of Physics, Tomsk State University,  
634050 Tomsk, Russia.}, Malte C. Tichy  
\vspace*{1.2\baselineskip}}

\affiliation{Institut f\"ur Theoretische Physik,
Universit\"at T\"ubingen,
\\ Auf der Morgenstelle 14, D--72076 T\"ubingen, Germany
\vspace*{0.3\baselineskip}\\}

\date{\today}

\begin{abstract} 

We study radial excitations of pseudoscalar and vector $q\bar q$ mesons 
within a chiral approach. We derive a general form for a chiral Lagrangian 
describing processes involving excited pseudoscalar and vector mesons.
The parameters of the chiral Lagrangian are fitted using data and 
previous calculations in the framework of the $^3P_0$ model.  
Finite--width effects are examined and predictions for mesons previously 
not discussed are given. Available experimental data is analyzed whenever 
possible. Possible hints for exotic mesons and open interpretation issues 
are discussed.

\end{abstract} 

\pacs{12.39.Fe, 13.25.Jx, 14.40.Cs} 

\keywords{chiral Lagrangian, pseudoscalar mesons, vector mesons, 
excited states} 

\maketitle

\newpage 

\section{Introduction}
The identification of non--$q\bar q$ mesons is a fascinating topic, 
being deeply connected to the nontrivial relationship between the 
QCD Lagrangian and the physical, observable meson states. 
Therefore, great experimental and theoretical efforts are being 
deployed in order to identify glueballs, hybrid mesons and others 
(for a recent review see e.g.~\cite{Klempt:2007cp,Mathieu:2008me}). 
However, exotic mesons may also possess nonexotic quantum numbers and, 
{\it a priori}, might not be distinguishable from ordinary $q \bar q$ states. 
The understanding of the decay pattern of excited pseudoscalar 
and vector mesons is therefore essential in order to identify 
exotic candidates, such as glueballs, hybrids, or meson molecules 
which carry ordinary quantum numbers. As no systematic theory exists 
in this nonperturbative regime, one has to rely on models in order 
to describe the decays. 

The spectrum and the decay pattern of excited mesons have been under 
study for many years. Descriptions of $q \bar q$--meson masses and 
decay properties have been done starting in the early eighties. 
In~\cite{Gerasimov:1981gj}, the radial excitations of light mesons have 
been considered in a phenomenological quark model, and the issue of the 
pseudoscalar isoscalars has already been adressed in~\cite{Gerasimov:1986ee}. 
The decay rates found will later be included in our discussion. 
In Ref.~\cite{Godfrey:1985xj} mesons (from the pion to the upsilon) have 
been considered in a relativized quark model. Meson masses and couplings 
have been calculated in a unified framework --- soft QCD. 
Decay amplitudes have been computed within the elementary emission model. 

The properties of vector--meson excitations were recently studied, e.g. 
the $\rho$--like mesons in~\cite{Surovtsev:2008zz}. On the interpretation 
of states on the experimental side, there is still some on-going debate. 
A recent analysis is given in~\cite{Achasov:2000hh,Achasov:2002uj}. 
The problem of excited scalar and tensor mesons has been under heavy 
investigation as well. The assignment of the structure of $f_0$ and 
$f_2$ (possibly with admixture of a glueball with corresponding quantum 
numbers) is an open issue, studied e.g. in~\cite{Surovtsev:2007nc}. 

In a series of 
papers~\cite{Volkov:1996br,Volkov:1996fk,Volkov:1999yi,Volkov:1997dd,
Volkov:1999xf,Volkov:2000ry} the masses of excited scalar, pseudoscalar 
and vector--meson states (the first excitation) and their respective 
couplings are studied in a chiral Lagrangian with form factors derived  
within the Nambu-Jona-Lasinio (NJL) quark model. 
Chiral symmetry breaking is described by the gap equation. 
Also, experimental candidates for the theoretical states are identified 
and the main strong decay widths are calculated. These will be confronted with
our results. Another approach to excited mesons has been performed
in Ref.~\cite{Holl:2004fr,Holl:2005vu}, using the Bethe--Salpeter equation, 
yielding masses and amplitudes of bound states in given $J^P$ channels. 

The $^3P_0$ model was used in several works to compute strong decays 
and serves today as the main reference for calculations of decay widths.
Decay properties for a wide mass range of mesons have been analyzed 
in~\cite{Barnes:1996ff,Barnes:2002mu,Barnes:2005pb}. Previously this 
model has been applied to the vector radial excitations 
in~\cite{Busetto:1982qz}. 

The most convenient language for the treatment of light hadrons at small 
energies was elaborated in the context of the chiral perturbation theory 
(ChPT)~\cite{Weinberg:1978kz,Gasser:1983yg,Gasser:1984gg,Ecker:1988te},  
the effective low--energy theory of the strong interaction. 
The goal of the present paper is the tree--level study of the decays of 
excited pseudoscalar and vector mesons within a chiral approach 
motivated by ChPT. We derive a general form of the chiral Lagrangian 
describing processes involving excited pseudoscalar and vector mesons. 
Parameters of the Lagrangian are fitted using data and 
previous calculations in the framework of the $^3P_0$ model.  
This work is continuation of a series of 
papers~\cite{Giacosa:2005bw,Giacosa:2005zt,Giacosa:2005qr}, where we 
analyzed the decays of scalar (including mixing with the scalar glueball) 
and tensor mesons (including mixing with the tensor glueball) using 
a phenomenological chiral approach. Here, we consider the strong decays 
of the first and second radial excitations in the pseudoscalar channel 
and the first excitation in the vector channel. Note that we do not 
consider mixing or additional components such as glueballs, yet. 
Since the experimental data is still very poor in this regime, a fit to the 
data would not be very useful. Whenever values are available, 
comparison will be done and evaluated.

The $^3P_0$ model has been used to predict many decay widths of mesons 
within a large mass range, from the GeV 
scale~\cite{Barnes:1996ff,Barnes:2002mu} up to 
charmonia~\cite{Barnes:2005pb}. Its phenomenological success has made 
it a popular tool for the estimation of decay widths. The model describes 
the strong open flavor decays of mesons as a $q \bar q$--pair production 
process. The two quarks of the produced pair separate to form the final 
mesons together with the two quarks of the initial meson. 
The $q \bar q$ pair is assumed to be produced in a $J^{PC}=0^{++}$ state 
corresponding to vacuum quantum numbers. This is the main assumption of 
the model. Simple harmonic oscillator functions are used for the incoming 
and outgoing particles, the decay amplitudes then have analytical 
expressions. Predictions obtained within the model are of a rather 
qualitative type, but have been shown to agree roughly with experiment, 
although large deviations can occur.

For many of the meson masses assumed 
in~\cite{Barnes:1996ff,Barnes:2002mu,Barnes:2005pb} better experimental 
data has become available now or different candidates for the theoretical 
states are being currently discussed. Therefore, in addition to the fit of 
our parameters, we comment the inclusion of finite--width effects and 
the impact of mass changes on the resulting decay widths. Different 
scenarios are analyzed in order to identify excited pseudoscalar and 
vector mesons.

In the present paper, we proceed as follows. In Sec.~\ref{sec:model}, 
we present the chiral Lagrangian which will be used and the 
generic expressions for the decay widths. The fit to the $^3P_0$ model 
within the different channels will be performed in the following section and 
we will discuss the consequences of available experimental data on the 
decays. We summarize our results in Sec.~\ref{sec:end} where 
we also discuss further possible applications. 

\section{Chiral Lagrangian for excited pseudoscalar 
and vector mesons}\label{sec:model}

The lowest--order chiral Lagrangian describing decays 
$P^\ast \rightarrow V P(V)$ and $V^\ast \rightarrow  P P(V)$
of excited vector $V^\ast$, pseudoscalar $P^\ast$ (first radial 
excitation), and $P^{\ast\ast}$ (second radial excitation) mesons 
involving also their ground states (pseudoscalars $P$ and vectors $V$) 
is motivated by 
ChPT~\cite{Weinberg:1978kz,Gasser:1983yg,Gasser:1984gg,Ecker:1988te},  
and reads as  
\eq\label{L_eff} 
{\cal L} &=&  \frac{F^{2}}{4} \left\la D_{\mu }U\,D^{\mu} 
U^{\dagger }+\chi _{+} \right\ra + {\cal L}_{\mathrm{mix}}^{P} 
- \frac{1}{2} \sum\limits_{{\cal V} = V, V^\ast} 
\left\la D^\rho {\cal V}_{\rho\nu} D_\mu {\cal V}^{\mu\nu} 
- \frac{1}{2} M_{\cal V}^2 {\cal V}_{\mu\nu} {\cal V}^{\mu\nu} 
\right\ra \nonumber\\  
&+& \frac{1}{2}  \sum\limits_{{\cal P} = P^\ast, P^{\ast\ast}}  
\left\la D^\mu {\cal P} D_\mu {\cal P} - M_{{\cal P}}^2 {\cal P}^2 \right\ra 
\nonumber\\[1mm]  
&+&  c_{_{P^\ast PV}} \, 
\left\la {\cal V}_{\mu\nu}  \, \left[ u^\mu, D^\nu {\cal P}^\ast 
\right] \right\ra + c_{_{P^\ast VV}} \, 
\left\la \epsilon^{\mu\nu\alpha\beta} \, {\cal P}^\ast  
\, \left\{ {\cal V}_{\mu\nu}, {\cal V}_{\alpha\beta} \right\} \right\ra 
+ c_{_{P^{\ast\ast} PV}} \, 
\left\la {\cal V}_{\mu\nu}\left[ u^\mu, D^\nu {\cal P}^{\ast\ast} 
\right] \right\ra  \nonumber\\[3mm] 
&+& c_{_{P^{\ast\ast} VV}} \left\la \epsilon^{\mu\nu\alpha\beta} 
{\cal P}^{\ast\ast} \, \left\{ {\cal V}_{\mu\nu}, 
{\cal V}_{\alpha\beta} \right\} \right\ra 
+c_{_{V^\ast PP}} \, \left\la {\cal V}^\ast_{\mu\nu} 
\left[ u^\mu , u^\nu  \right]  \right\ra + c_{_{V^\ast VP}} \, 
\left\la \epsilon^{\mu\nu\alpha\beta} \, U \, 
\left\{ {\cal V}_{\mu\nu}, {\cal V}^\ast_{\alpha\beta}\right\} 
 \right\ra \,. \en 
Here we use the following notation: The symbols $\la \, \cdots \, \ra$, 
$[ \, \cdots \, ]$ and $\{ \, \cdots \, \}$ occurring in Eq.~(\ref{L_eff}) 
denote the trace over flavor matrices, commutator, and anticommutator, 
respectively. $U=u^{2}=\exp (i P\sqrt{2}/F)$ is the chiral field collecting 
ground state pseudoscalar fields in the exponential parametrization with 
\eq 
P = 
\left(\begin{array}{ccc} 
  \displaystyle\frac{\pi^0}{\sqrt{2}} 
+ \displaystyle\frac{\eta_8}{\sqrt{6}}  & \pi^+ & K^+\\
\pi^-& - \displaystyle\frac{\pi^0}{\sqrt{2}} 
+ \displaystyle\frac{\eta_8}{\sqrt{6}} & K^0\\
K^-& \overline{K^0} & 
- \displaystyle\frac{2}{\sqrt{6}} \eta_8 
\end{array}\right)  \,. 
\en
$D_{\mu }$ denotes the chiral and gauge--invariant derivative, 
\hspace*{0.2cm} $u_{\mu}=iu^{\dagger }D_{\mu }Uu^{\dagger }$ is 
the chiral field, 
$\chi _{\pm}=u^{\dagger }\chi u^{\dagger } \pm u \chi^{\dagger }u, 
\hspace*{0.2cm}\chi=2B(s+ip),\,\,\,s=\mathcal{M} + \cdots \,$ and 
$\mathcal{M}=\mathrm{diag}\{\hat{m},\hat{m},m_{s}\}$ is the 
current quark mass (we restrict to the isospin symmetry limit with 
$m_{u}=m_{d}=\hat{m}$); $B$ is the quark vacuum condensate parameter 
and $F$ the pion decay constant. The matrices ${\cal V}$ and 
${\cal V}^\ast$ represent the nonets of ground state 
$\{ \rho^\pm, \rho^0, K^{\ast \, \pm}, K^{\ast \, 0}, 
\overline K^{\ast \, 0}, \phi \}$ and first radially excited 
$\{ \rho^\pm(1450)$, $\rho^0(1450)$, $\omega(1420)$, 
$K^{\ast \, \pm}(1680)$, $K^{\ast \, 0}(1680)$, 
$\overline K^{\ast \, 0}(1680)$, $\phi(1680)\}$ vector mesons 
in tensorial representation~\cite{Gasser:1983yg,Ecker:1988te}. 
 ${\cal P}^\ast$ and ${\cal P}^{\ast\ast}$ denote the nonets of 
the first $\{ \pi^\pm(1300)$, $\pi^0(1300)$, $K^\pm(1460)$,  
$K^0(1460)$, $\overline{K^0}(1460)$, $\eta(1295)$, $\eta(1475) \}$ 
and second $\{ \pi^\pm(1800)$, $\pi^0(1800)$, $K^\pm(1830)$,  
$K^0(1830)$, $\overline{K^0}(1830)$, $\eta(1760)$, $\eta(2225) \}$ 
radially excited pseudoscalar meson fields (see explicit form of ${\cal V}$, 
${\cal V}^\ast$, ${\cal P}^\ast$ and ${\cal P}^{\ast\ast}$ 
in Appendix A).  The constants $c_{_{P^\ast PV}}$, $c_{_{P^\ast VV}}$, 
$c_{_{P^{\ast\ast} PV}}$, $c_{_{P^{\ast\ast} VV}}$,  
$c_{_{V^\ast PP}}$, $c_{_{V^\ast VP}}$ define the couplings 
between the corresponding types of mesons, respectively. 
Following~\cite{Ecker:1988te} we encode in $\mathcal{L}_{\mathrm{mix}}^{P}$
an additional contribution to the mass of the $\eta_0$
(due to the axial anomaly)
and the $\eta_0$ -- $\eta_8$ mixing term: 
\eq\label{L_p_mix} 
\mathcal{L}_{\mathrm{mix}}^{P} = 
- \frac{1}{2}\gamma_{P}  \eta_0^2   
- z_{P} \eta_0 \eta_8 \,,
\en
(the parameters $\gamma _{P}$ and $z_{P}$ are in turn related to the
parameters $M_{\eta _{1}}$ and $\widetilde{d}_{m}$ of~\cite{Ecker:1988te}).
The physical, diagonal states $\eta $ and $\eta ^{\prime }$ are given by 
\eq 
\eta_0 \, = \, \eta^\prime \, \cos\theta_P \, - \, \eta \, 
\sin\theta_P\,, \hspace*{0.25cm} 
\eta_8 \, = \, \eta^\prime \, \sin\theta_P \, 
+ \, \eta \, \cos\theta_P \,, 
\en 
where $\theta_{P}$ is the pseudoscalar mixing angle. 
We follow the standard procedure~\cite{Ecker:1988te,Cirigliano:2003yq,%
Kawarabayashi:1980dp,Venugopal:1998fq,Giacosa:2005zt} and
diagonalize the corresponding $\eta_0$ -- $\eta_8$ mass matrix to obtain
the masses of $\eta $ and $\eta^{\prime }$. By using $M_\pi=139.57$~MeV, 
$M_K=493.677$~MeV (the physical charged pion and kaon masses), 
$M_\eta=547.75$~MeV and $M_{\eta ^{\prime }}=957.78$~MeV, the mixing angle 
is determined as $\theta_{P}=-9.95^{\circ }$, which corresponds to the
tree--level result (see details in Ref.~\cite{Venugopal:1998fq}).
Correspondingly, one finds $M_{\eta_0}=948.10$ MeV  
and $z_{P}=-0.105$ GeV $^{2}.$
Higher order corrections in ChPT cause a doubling of the absolute value of
the pseudoscalar mixing angle~\cite{Venugopal:1998fq}; we restrict our work 
to the tree--level evaluation, we therefore consistently use the
corresponding tree--level result of $\theta _{P}=-9.95^{\circ }$.
In the present approach we do not include the neutral pion when considering
mixing in the pseudoscalar sector, because we work in the isospin limit. 
For all pseudoscalar
mesons we use the unified leptonic decay constant $F$, which is identified
with the pion decay constant $F=F_{\pi }=92.4$~MeV. A more accurate analysis
including higher orders should involve the individual couplings 
of the pseudoscalar mesons (for a detailed discussion 
see Refs.~\cite{Gasser:1984gg}).
For the radial excitations and the ground state vector mesons, 
we assume ideal mixing, so that the flavor content of the 
excited $\phi$ and $\eta_{ss}$ is completely given by $s \overline s$, 
the content of $\omega$ and $\eta_{nn}$ by
$n \overline n =(u \bar u + d \bar d)/\sqrt{2}$.

The width for a generic two--particle decay $A \to B \, C$ 
is given by
\eq \Gamma(A \rightarrow B  \, C) = 
f \frac{\lambda^{3/2}(m_A^2, m_B^2, m_C^2)}{\pi m_A^3 } 
\left| M \right|^2 
\en 
where $\lambda(x,y,z) = x^2 + y^2 + z^2 - 2 xy - 2xz -2yz$  
is the K\"all\'en function; 
$M$ denotes the amplitude squared including the parameter of 
the decay strength; $f$ denotes an additional factor which is 
$\frac 1 2$ for $P^\ast\rightarrow VV$, $\frac 1 6$ for 
$V^\ast \rightarrow PP$, $\frac 1 {64}$ for $P^\ast\rightarrow P V$ 
and $ \frac 1 {192}$ for $V^\ast \rightarrow VP$. The average over 
polarization has already been included, a factor of $\frac 1{2\!}$ 
will have to be added when considering decays with two identical 
particles in the final state. The decay amplitudes $M$ in tree level 
are given in the Tables in Appendix \ref{app:tables}.

In the following analysis of the decay widths, we will often encounter 
decay modes which are kinematically strongly suppressed or forbidden,
when using the central mass values. Because of the finite width of the 
decaying or the produced particles, these 
decays may however be enhanced considerably. We include this effect in 
our calculations by approximately taking into account the mass distribution 
of a meson with a certain width in the following way
\eq 
\Gamma_{\rm full}(A\rightarrow B \, C) = 
\int \Gamma(A \rightarrow B \, C)_{m_A=m} 
f(m) \, dm \,. \label{broadwidth}
\en 
The function $f(m)$ is the mass distribution of a resonance with central
mass $M$ and total width $\Gamma$ given by
\eq\label{fm_ansatz} 
f(m) = \left\{ 
\begin{array}{ll}  
0\,, & \hspace*{1cm} m < A_{\rm thr} \\ 
\displaystyle\frac1{ 4 A_0 } \frac{\Gamma^2}{(m-M)^2+\frac 1 4 \Gamma^2} 
\cdot \frac{m-A_{\rm thr}}{M-\Gamma-A_{\rm thr}} \,, 
& \hspace*{1cm}  A_{\rm thr} < m < M-\Gamma \\ 
\displaystyle\frac1{ 4 A_0 } 
\frac{\Gamma^2}{(m-M)^2+\frac 1 4 \Gamma^2} \,, 
& \hspace*{1cm} M - \Gamma < m < M + \Gamma \\ 
\displaystyle\frac1{ 4 A_0 } \displaystyle\frac{\Gamma^2}{(m-M)^2 
+ \displaystyle\frac 1 4 \Gamma^2} \cdot 
\displaystyle\frac{M+3\Gamma - m }{2 \Gamma}\,, 
& \hspace*{1cm}  M +  \Gamma < m < M + 3 \Gamma \\ 
0 \,, & \hspace*{1cm} M + 3 \Gamma < m ~,
\end{array} \right. 
\en  
where $A_{\rm thr}$ denotes the threshold and $A_0$ is a normalization 
constant such that 
\eq 
\int f(m) \, dm = 1 \,. 
\en
Also for broad final states an analogous integration, as suggested in
Eq.~(\ref{broadwidth}), will be performed.  
Note, the spectral function $f(m)$ is taken as a Breit--Wigner form
where the low-mass tail is modified to introduce a proper threshold
cutoff $A_{\rm thr}$.
For the high mass tail an additional regularization is introduced
to keep the distribution for a finite range of mass values.
Such parametrizations of the spectral function were originally used
and tested in the study of 
proton-antiproton annihilations into mesons~\cite{Maruyama:1980rh},
where finite size effects
can also play a relevant role.
We will see that in many cases the inclusion of finite--width effects
is important, especially when the central masses of the final state mesons
lie near threshold.

\section{Fit of the decay strengths and results}
 
In the present approach, relative rates are parameter free predictions, when
staying within given nonets both in the initial and final state. To also obtain
absolute values for the decay widths we use predictions of the 
$^3P_0$ model~\cite{Barnes:1996ff,Barnes:2002mu} to set a scale for the 
corresponding coupling constants. In particular, for a given set of decay 
modes we perform a $\chi^2$ fit to the predictions 
of~\cite{Barnes:1996ff,Barnes:2002mu} and also give an explicit comparison.

For experimentally measured mesons we use the notation of 
the Particle Data Group (PDG); for the 
theoretical states we use $\pi^\ast, \eta^\ast$, etc. The fits to the
$^3P_0$ model are performed 
without inclusion of finite--width effects since the authors 
in~\cite{Barnes:1996ff,Barnes:2002mu} do not include these either. 

In our model, we do not include the radial excitations in the final state
channels since this 
would increase the number of parameters too much. Since the radial excitations 
are not identified unambiguously yet, this would be a further source of 
uncertainty. However, the decay into excited mesons becomes only relevant 
when looking at the second excitation of the vector mesons, therefore this
possible fit is not presently discussed. 

\subsection{First radial pseudoscalar excitation}

\subsubsection{Experimental situation}
The common interpretation (see e.g. Ref.~\cite{Anisovich:2002us}) 
of the first radial excitation of the pseudoscalar mesons in terms of the 
measured resonances (as given by PDG~\cite{Amsler:2008zz}) 
is the following : 
\eq 
\left(\begin{array}{c}
 \pi^\ast \\
 \eta_{nn}^\ast \\
 \eta_{ss}^\ast \\
 K^\ast 
 \end{array}\right) 
 &=&
\left(\begin{array}{c}
 \pi(1300) \\
 \eta(1295) \\
 \eta(1405), \eta(1475) \\
 K(1460) 
 \end{array}\right) ~. 
\en
Because of the mass degeneracy of the $\pi(1300)$ with the $\eta(1295)$, 
the interpretation of the latter state as dominantly composed of 
$n \overline n$ is strengthened and points to a rather small mixing angle 
with any other higher state. The question whether the $\eta(1405)$ or the 
$\eta(1475)$ is a quarkonium or an exotic state remains open despite 
several recent attempts to disentangle this problem, possibly connected 
to the existence of a glueball 
(see e.g. Ref.~\cite{Gerasimov:2007sb}, an overview on the search of 
the pseudoscalar glueball is given in~\cite{Masoni:2006rz}). 
We will show in the framework of our analysis how to possibly clarify
this situation with further measurements. 

The $K(1460)$ is not yet accepted by PDG, but it is the only candidate 
with the right quantum numbers for an excited kaon in this mass range. 
Its mass, which has been measured to be between 1400 and 1460 MeV, 
is compatible with a kaon state about 150 MeV higher than the $\pi(1300)$ 
due to the $s$ quark. For the $K(1460)$ we will study the effect of different 
masses and finite--width effects. Besides the puzzling situation concerning 
the supernumerous state, the identification in this sector is rather clear. 

\subsubsection{Decays and fit}

In Table \ref{table:p1}, we give our fit (chiral approach) to the 
$^3P_0$ model~\cite{Barnes:1996ff,Barnes:2002mu}, indicating latter results 
and the predictions of other theoretical 
approaches~\cite{Gerasimov:1981gj,Volkov:1997dd} for the decays of the 
first radial pseudoscalar excitation into pseudoscalar and vector mesons. 
The coupling constant results are 
\eq 
c_{_{P^\ast PV}} = 4.95 \ {\rm GeV}^{-1} ~, \label{coupPPV}
\en 
where the value does not depend on which of the $\eta_{ss}$ candidates is 
used.
Note that chiral and $^3P_0$ approaches have an agreement in 
the results for the $P^\ast \to PV$ decay widths, which are 
largely determined by quantum numbers, phase space, and flavor
symmetry which are common to both models. In addition, 
the $^3P_0$ model assumes identical wavefunctions across flavor 
multiplets, also in agreement with the chiral model.

\begin{table} 
\begin{center}
\caption{Theoretical results for the $P^\ast \rightarrow PV$ 
decay widths (in MeV).}\label{table:p1}
\begin{tabular}{|cc|c|c|c|c|} \hline
 Decay mode &  
& Gerasimov {\it et al.} \cite{Gerasimov:1981gj} 
& Volkov {\it et al.} \cite{Volkov:1997dd} 				
& Barnes {\it et al.} \cite{Barnes:1996ff,Barnes:2002mu} 
& Chiral approach  \\ 
\hline 
$\pi(1300)$ & $\rightarrow \pi \rho $ & $630 \pm 160$     
& 220 & $ 209 $ & $ 257$  \\ \hline
$ \eta_{ss}(1415) $ & $\rightarrow K \overline{K^\ast}$  
&-& $\sim 0$ & $ 11 $ & $ 10$ \\
$ \eta_{ss}(1500) $ & $\rightarrow K \overline{K^\ast}$  
&-& $\sim 0$ & $ 100 $ & $ 82$  \\ \hline
$ K(1460) $ 	      & $\rightarrow \rho K$
&-& 50 & $ 73 $ & $ 65$  \\ 
& $\rightarrow \omega K $ 				
&-&- & $ 23 $ & $ 20$  \\ 
& $\rightarrow \pi  K^\ast $ 				
&-& 100 & $ 101 $ & $ 127 $  \\ 
& $\rightarrow \eta K^\ast $ 				
&-&- & $ 3 $ & $ 2$  \\  
\hline
\end{tabular}
\end{center}
\end{table}

It is important to mention that in Ref.~\cite{Volkov:1997dd} the 
excited kaon mass was assumed as 1.3 GeV, which leads to a considerable
reduction in phase space.
Because of interference the approach in~\cite{Volkov:1997dd} 
predicts a very narrow width for the decay of 
$\eta_{ss}(1470) \rightarrow K \overline{K^\ast})$. 
In Ref.~\cite{Gerasimov:1981gj}, the decay width 
$\Gamma(\pi^\ast \rightarrow \pi \rho)$ was calculated using available 
data on $\Gamma(\rho^\ast \rightarrow \omega \pi)$ at that 
time, i.e. assuming a mass of 1.1 to 1.2 GeV. 

\subsubsection{The $\eta_{ss}$ excitation}

Barnes {\it et al.} assumed in \cite{Barnes:1996ff,Barnes:2002mu} 
that the $\eta_{ss}$ lies between 1.415 and 1.5 GeV. Presently, 
the experimental candidates for this meson are the $\eta(1405)$ and 
$\eta(1475)$. It is still unclear which one is to be considered supernumerous 
and what values the mixing angle may take. Using the coupling constant 
of (\ref{coupPPV}), we give the decay widths for these masses including and, 
in brackets $(\cdots)$, excluding finite--width effects
\eq 
\Gamma(\eta_{ss}(1476) \rightarrow K \overline{K^\ast} )
&=& 67 \ \mbox{MeV} \ (57 \ \mbox{MeV}) \,, \\
\Gamma(\eta_{ss}(1410) \rightarrow K \overline{K^\ast} )  
&=& 13 \ \mbox{MeV} \ (7 \ \mbox{MeV}) \,. 
\en 
The decay width strongly depends on the mass value of the state 
and on finite--width effects, especially for the 
lower state. If we compare this value with the full width, the result 
seems to point to an $s\bar s$ interpretation of the $\eta(1475)$: 
The full width of the experimental state $\eta(1475)$ is 
$87 \, \mbox{MeV}$ and the decay mode $K \overline K \pi$ fed by 
$K \overline{K^\ast}$ is dominant. 
The absence of the $K \overline{K^\ast}$ decay mode for the 
experimental candidate $\eta(1405)$ seems to point to the conclusion that 
its $s\bar s$ component is small. However, the expected decay width would 
be small anyway due to the strong kinematical suppression. It is important 
to stress the fact that the nonobservation of this mode is not necessarily
clear evidence for the absence of a large $s \bar s$ admixture. 

Further measurements of the decay to $K\overline{K^\ast}$ would be useful in 
this case to better identify the $\eta_{ss}$ state in this region and to sort 
out the supernumerous state. A measurement of the 
direct three--body decays $\eta(1405)\rightarrow \eta \pi \pi, K \bar K \pi$,
to be discriminated from resonance--fed decays, could help to estimate the 
mixing angle in different mixing scenarios in this mass region, since a chiral 
approach could describe three--body decays as well. Unfortunately, 
the lack of experimental data makes a serious analysis impossible 
at the moment.

\subsubsection{Finite--width effects of $\eta (1295)$ and $\pi(1300)$}

The experimental state $\eta (1295)$ has a width of 
$55\pm 5$ MeV. The radially excited pion is even broader, 
$\Gamma(\pi(1300))\ge 200$ MeV. Now we include finite--width effects
to study the $K \overline{K^\ast}$ decay channel. 

We compute the $K \overline{K^\ast}$ decay width using 
expression (\ref{broadwidth}) and find 
\bea  
\Gamma(\pi(1300) \rightarrow K\overline{K^\ast})&=&10.37 \ \mbox{MeV} \,,\\ 
\Gamma(\eta(1295)\rightarrow K\overline{K^\ast})&=&0.07 \ \mbox{MeV} \,. 
\ena
For a $n \bar n$ $\eta$ state at 1.3 GeV the decay width into 
$K\overline{K^\ast}$ is negligible. The excited pion however should decay 
with a small finite width into $K K^\ast$. Note that the decay 
$\pi (1300) \to \pi \rho$ is absolutely dominant as stated by the theoretical 
predictions and seems to feed the full width of 200--600 MeV. Indeed, the 
decay into $K\overline{K^\ast}$ has not 
been seen so far. For the $\eta$ an upper bound of the $\pi \pi K$ width 
exists which is consistent with the present small estimate for this width.

\subsubsection{The kaon}

The experimental situation regarding the kaon is not very clear, there are 
two possible mass assignments (1.4 GeV and 1.46 GeV). The total width is 
large, around 250 MeV in both cases. In Table~\ref{table:kaon14} we give 
the decay widths using our approach including finite--width effects for both 
candidates.

\begin{table} 
\begin{center}
\caption{Decay width including finite--width effects 
for the kaon (in MeV).}\label{table:kaon14} 
\begin{tabular}{|cc|c|c|c|} \hline
 Decay mode & & $m$ = 1.4 GeV & $m$ = 1.46 GeV & Data~\cite{Amsler:2008zz} \\ 
\hline 
$ K $ & $\rightarrow \rho K $      & 73  & 96  & 34 \\
&       $\rightarrow \omega K $    & 21  & 30  & - \\
&       $\rightarrow \pi  K^\ast $ & 118 & 152 & 109 \\ 
&       $\rightarrow \eta K^\ast $ & 21  & 34  & - \\
&       $\rightarrow \phi K      $ & 8   & 13  & -\\ \hline
\end{tabular}
\end{center}
\end{table}

The decay modes $\eta K^\ast$ and $\phi K$ are 
considerably enhanced by finite--width effects. The observed decay 
width to $\rho K$ of around $34$ MeV (no error is given by PDG) 
is overestimated by a factor of 3 in the model prediction.
The decay width to $K^\ast \pi$ is measured as 109 MeV and lies in the 
range of our fit. A direct fit of the coupling constant to the data
would not allow an improved correspondence, since the ratio 
\eq 
\frac{\Gamma(K(1400)\rightarrow \rho K)}{\Gamma(K(1400)\rightarrow 
\pi K^\ast)}\sim 6 
\en 
is fixed in the present model, but experimentally it is currently found 
to be $\sim 0.3$. Even though the 
decay to $K^\ast(1430) \pi$ is strongly suppressed kinematically, 
the experimental value of 117~MeV is rather large. Further data on
decay modes involving the kaon would be useful to establish this 
state better, the experimental data used here is extracted from one single 
experiment and is not confirmed by PDG. Since its mass lies in a reasonable 
range with respect to the pion--nonet partner, no exotic scenario is evident 
here.

\subsection{Second radial pseudoscalar excitation}

The mass region of the second radially excited pseudoscalar mesons 
is interesting since it lies closer to the region in which lattice QCD 
predicts the pseudoscalar glueball with mass of about 
$2.3\pm 0.2$ GeV~\cite{Bali:1993fb}. 
Mixing might lower the mass considerably and affect the pattern in 
this mass region. So far, no supernumerous state has been observed in this 
region, but no clear candidate for $\eta_{ss}$ has been observed so 
far either.

\subsubsection{Experimental situation}

A possible interpretation of the second radial excitations would be as
follows: 
\eq 
\left(\begin{array}{c}
 \pi^{\ast\ast} \\
 \eta_{nn}^{\ast\ast} \\
 \eta_{ss}^{\ast\ast} \\
 K^{\ast\ast} 
 \end{array}\right) 
 &=&
\left(\begin{array}{c}
 \pi(1800) \\
 \eta(1760) \\
 ? \\
 K(1830) 
 \end{array}\right) ~.
\en
There is no clear candidate for $\eta_{ss}$. The next resonance with suitable 
quantum numbers would be the $\eta(2225)$ which lies considerably higher than 
expected and has been discussed as a glueball candidate. The candidate for 
the kaon has a rather low mass, being very close to the pion. 
This phenomenon can be understood by the following argument. 
The larger strange quark mass implies a smaller
excitation energy for kaons. Eventually the kaon--ike states
come closer to and become even lighter than the nonstrange
counterparts, typically around the second or third excitations.
The candidate for $\eta_{nn}$ lies somehow lower than the corresponding pion 
state, which might allude to a possible mixing scenario which we will discuss.

\subsubsection{The fit}

\begin{table}
\begin{center}
\caption{Decay widths of second radial pseudoscalar excitation 
to pseudoscalars and vectors (in MeV).}\label{table:p2pv}
\begin{tabular}{|cc|c|c|c|c|} \hline
 Decay mode & 
& Gerasimov {\it et al.} \cite{Gerasimov:1981gj} 
&Barnes {\it et al.} \cite{Barnes:1996ff,Barnes:2002mu} 
& Chiral approach (1) & Chiral approach (2) \\ 
\hline
$\pi(1800)$ & $\rightarrow \pi \rho $  & 32-50 & 31 & 54 & 77 \\
&$ \rightarrow K \overline{K^\ast}$ &-& 36 & 13 & 18\\ \hline
 $\eta_{nn}(1800)$& $\rightarrow K \overline{K^\ast}$ &-& 36 & 13 & 18 \\ 
\hline
 $\eta_{ss}(1950) $ & $ \rightarrow K \overline{K^\ast}$ & -& 53 & 43 & 61 \\ 
\hline 
 $ K(1830) $ 				
& $\rightarrow \rho K $      &-& 21 & 15 & 21 \\ 
& $\rightarrow \omega K $    &-& 7  & 5  & 7  \\ 
& $\rightarrow \phi K $      &-& 18 & 4  & 6  \\ 
& $\rightarrow \pi  K^\ast $ &-& 16 & 18 & 25 \\ 
& $\rightarrow \eta K^\ast $ &-& 27 & 9  & 12 \\  
\hline
\end{tabular}
\end{center}
\end{table}

\begin{table}
\begin{center}
\caption{Decay widths of second radial pseudoscalar 
excitation to two vectors (in MeV).\label{table:p2vv}}
\begin{tabular}{|cc|c|c|c|} \hline
 Decay mode & & Barnes {\it et al.}~\cite{Barnes:1996ff,Barnes:2002mu} 
& Chiral approach  \\ 
\hline
$\pi(1800)$ & $\rightarrow \rho \omega $ & 73 & 83 \\ \hline 
$\eta_{nn}(1800)$ &$ \rightarrow \rho \rho$ & 112 & 130 \\ 
&$  \rightarrow \omega \omega$ & 36 & 40 \\ \hline   
$\eta_{ss}(1950) $ & $ \rightarrow K^\ast \overline{K^\ast}$ & 67 & 55 \\ 
\hline
 $ K(1830) $ & $\rightarrow \rho K^\ast $ & 45 & 36 \\ 
& $\rightarrow \omega K^\ast $ & 14 & 11 \\ \hline
\end{tabular}
\end{center}
\end{table}

In the chiral approach, these widths are related to the couplings 
$c_{P^{\ast\ast} PV}$ and to $c_{P^{\ast\ast} VV}$, 
the best fit to the results of the $^3P_0$ model gives 
\eq 
& &c_{_{P^{\ast\ast} PV}} = 0.94589 \ {\rm GeV}^{-1} \,, \nonumber\\
& &c_{_{P^{\ast\ast} VV}} = 0.29759 \ {\rm GeV}^{-1} \,. 
\en 
Our results in comparison with other theoretical predictions are given 
in Tables~\ref{table:p2pv}, \ref{table:p2vv} and \ref{table:etapi}. 
 
\subsubsection{General discussion}

The authors in Ref.~\cite{Volkov:1997dd} do not give predictions for 
the decays of the second radial excitation of the pseudoscalar~($3^1S_0$). 
In the vector--vector channel, the chiral approach reproduces the decay 
pattern rather well. The order of magnitude of the decay widths are 
equivalent, the general pattern is the same. 
In the vector--pseudoscalar channel strong deviations between the results of
the $^3P_0$ model and the present predictions are observed.
In Ref.~\cite{Gerasimov:1981gj}, the fact that
\eq 
\Gamma\left(\pi^{\ast\ast} \rightarrow \pi \rho\right) 
\ll \Gamma\left(\pi^{\ast} \rightarrow \pi \rho\right) 
\en
is well explained. The reason for this naively unexpected behavior 
can be traced to the node structure of the radial wave functions of the
excited pion states.
The prediction was indeed 
confirmed experimentally and solidifies the present interpretation 
of $\pi^\ast, \pi^{\ast\ast}$ as radial excitations.
This behavior is also present in the
basis for our fit (1), that is the predictions of the $^3P_0$ model
in~\cite{Barnes:1996ff,Barnes:2002mu}. Here, the suppression of the 
$\pi^{\ast\ast} \rightarrow \pi \rho $ mode is reflected
by a rather small coupling constant in order to fit the
small width $\Gamma(\pi(1800) \rightarrow \pi \rho)=31$ MeV found by
Barnes {\it et al.}~\cite{Barnes:1996ff,Barnes:2002mu}.
At the same time, the additional decay width
for $K \overline{K^\ast}$ is consequently smaller in the chiral approach. 
Our second fit (2) does not include the decay width to $\pi \rho$ and leads 
to a larger coupling constant: 
\eq 
c_{_{P^{\ast\ast} PV}} = 1.1295 \ {\rm GeV}^{-1} \,. 
\en  
The relation $\Gamma(\eta(1800) \rightarrow K \overline K^\ast) = 
\Gamma(\pi(1800) \rightarrow K \overline K^\ast)$ also emerges naturally 
in the chiral approach, but the authors of 
Ref.~\cite{Barnes:1996ff,Barnes:2002mu} 
find $\Gamma(\pi(1800) \rightarrow \rho \pi) 
\le \Gamma(\pi(1800) \rightarrow K \overline{K^\ast})$. In our approach, 
independent of the choice of the coupling, we have
\eq 
\frac{\Gamma(\pi(1800) \rightarrow \rho \pi)}{\Gamma(\pi(1800) 
\rightarrow K \overline{K^\ast})} \ = \ 
2 \, \frac{\lambda(m_{\pi^{\ast\ast}}^2,m_\pi^2,m_\rho^2)^{3/2}}
{\lambda(m_{\pi^{\ast\ast}}^2,m_K^2,m_{\overline{K^\ast}}^2)^{3/2}} 
= 4.35 \,. 
\en 

\subsubsection{$\eta_{nn}$ and $\pi$} 

We can use our results to identify the theoretical states with experimental 
candidates. The experimental states for the second radial excitation of 
$\eta_{nn}$ and $\pi$ are easily found: They are the $\eta(1760)$ with mass 
$1756 \pm 9$~MeV and total width $96 \pm 70$~MeV. The approximately 
mass--degenerate pion--partner would be the $\pi(1800)$ with mass 
$1816 \pm 14$~MeV and width $208 \pm 12$~MeV. 

The predictions for the partial decay widths 
of $\eta(1800)$, given in~\cite{Barnes:1996ff,Barnes:2002mu} with a mass of 
1.8 GeV, are too high. Because of the lower mass of the experimental candidate 
(1.76 GeV) they have to be corrected. 
Since the decays to $\rho\rho$ and $\omega \omega$ are close to threshold, 
the small change in mass changes the decay pattern considerably. 
The large width of the initial state has to be considered as well to check 
whether we can expect a sizable decay to $K^\ast \overline{K^\ast}$ which 
lies kinematically near threshold. For the experimental candidates, we find 
the decays listed in Table \ref{table:etapi}.

\begin{table} 
\begin{center}
\caption{Prediction of decay widths for 
experimental candidates (in MeV).}\label{table:etapi}
\begin{tabular}{|cc|c|c|} \hline
 Decay mode & & Chiral approach (zero width) & Chiral approach 
(finite width) \\ 
\hline
 $\eta(1760)$& $\rightarrow K \overline{K^\ast}$ & 10 & 10 \\ 
&$ \rightarrow \rho \rho$ 		         & 95 & 94 \\ 
&$  \rightarrow \omega \omega$ 			 & 29 & 29 \\ 
 &$ \rightarrow K^\ast K^\ast$ & 0 & 2 \\ \hline 
 $\pi(1800)$& $\rightarrow K^\ast \overline{K^\ast}$ & 2 & 26 \\ 
& $\rightarrow \rho \omega $ & 90 & 217 \\  
& $\rightarrow K K^\ast  $ & 13 & 13 \\  
& $\rightarrow \pi \rho  $ & 56 & 55\\   \hline
\end{tabular}
\end{center}
\end{table}

The $\eta (1760) \to  K^\ast \overline{ K^\ast}$ decay mode is essentially 
suppressed, for the $\pi (1800)$ however, we can indeed expect to see this 
decay. The enhancement of the $\rho\omega$ channel for the $\pi$ is very 
strong, too. For the $\pi (1800)$ the situation is presently rather unclear, 
since the mode $\pi\rho$ has not been observed yet. Also, the dominant decay 
channel $\rho\omega$, which feeds the 5$\pi$ final state, has not been 
detected yet.

\subsubsection{Identification of $\eta_{ss}$}

The $\eta_{ss}$ meson is assumed to lie around 2000 MeV 
(the approach in~\cite{Barnes:1996ff,Barnes:2002mu} uses a mass of
1950 MeV), close to the $\phi \phi$ threshold. 
Predictions for this decay mode will therefore 
depend strongly on the mass and width of the $\eta_{ss}(\approx 2000)$.

The identification of $\eta_{ss}$ is difficult: No pseudoscalar 
isoscalar state has been observed so far near 2000 MeV. 
The next candidate would be the resonance $\eta(2225)$, 
which, however, lies much higher than one would naively expect. 
It is a broad state (the width is about 150 \mbox{MeV}), 
hence finite--width effects can be important.

\subsubsection{Mixing scenario}

As the $\eta_{nn}$ candidate lies somewhat lower, one can analyze 
a mixing scenario of a bare $\eta_{nn}$ around 1812 MeV and a bare 
$\eta_{ss}$ at a mass lower than 2225 MeV. One finds
\eq  
\left(\begin{array}{c} 
\eta(1760) \\ 
\eta(2225) 
\end{array} \right) = 
\left(\begin{array}{cc} 
0.95 & 0.32 \\ 
-0.32 & 0.95
 \end{array} \right) 
\left( \begin{array}{c} \eta_{nn}(1812) \\ 
\eta_{ss}(2183) 
\end{array}\right) \,  , 
\en 
which corresponds to a mixing angle of $\alpha = 0.32$. 
This scenario may help to explain the large mass of the 
$\eta_{ss}$ candidate and the low mass of $\eta_{nn}$. 
However, no strong mixing dynamics is presently known which would induce
such mass shifts.
If the pseudoscalar glueball is 
present in this mass region, mixing of this state with 
$\eta_{nn}$ and $\eta_{ss}$ might also result in a similar mass shift.

Let us assume that the $\eta(2225)$ state is indeed mainly $\eta_{ss}$. 
Its decay pattern is listed in Table~\ref{table:eta222}. The total 
width of the experimental candidate with 
$\Gamma =150^{+300}_{-60}\pm60$ MeV would be largely overestimated in 
the fit of the coupling constant to the $^3P_0$ data. The decay rates 
have not been measured yet. 

\begin{table} 
\begin{center}
\caption{Predictions for the partial decay widths of $\eta_{ss}(2220)$ 
(in MeV)}\label{table:eta222} 
\begin{tabular}{|cc|c|c|} \hline
Decay mode & & Chiral approach (zero width) & Chiral approach (finite width) 
\\ 
\hline 
$\eta_{s s}(2220)$ & $\rightarrow K \overline{K^\ast} $ & 79 & 78 \\ 
 & $\rightarrow \phi \phi $  & 40 & 44 \\
 & $\rightarrow K^\ast \overline {K^\ast} $ & 265 & 264 \\ 
\hline 
\end{tabular}
\end{center}
\end{table}

An important question would be to clarify whether this state is the 
third radial excitation of the $\eta_{nn}$ configuration.
A mass--degenerate pion triplet 
with the same mass would of course favor this interpretation. 
For this purpose we analyze the decay ratios for both an $\eta_{ss}(2225)$ 
and $\eta_{nn}(2225)$, where we assume the width to be 150 MeV
(and the threshold set to 1 GeV):
\eq 
\Gamma(\eta_{ss} \rightarrow \phi \phi) : 
\Gamma(K^\ast \bar{K^\ast}) & = & 1:6 \,,\\
\Gamma(\eta_{nn} \rightarrow \omega \omega) :  
\Gamma(K^\ast \bar{K^\ast}):\Gamma( \rho \rho)  & = & 1:1.20:4.62 \,. 
\en 
While the $K^\ast \bar{K^\ast}$ decay mode is the strongest $VV$ mode 
for the $\eta_{ss}$ state, for $\eta_{nn}$ we should rather expect the 
$\rho\rho$ as dominant decay mode.
A determination of the $K^\ast \bar{K^\ast}$ channel would help in further
clarification of this state.

The situation for the $\eta (2225)$ is not satisfactory. The knowledge of 
the decay widths might clarify the nature of the $\eta(2225)$ and give 
evidence for or against its interpretation as mainly $\eta_{ss}$.

\subsection{First radial vector excitations}

\subsubsection{Experimental situation}
The experimental candidates for the first radial vector excitations are 
the following:
\eq 
\left(\begin{array}{c}
 \rho^\ast \\
 \omega^\ast \\
 \phi^\ast \\
 K^\ast  
 \end{array}\right) 
 &=&
\left(\begin{array}{c}
 \rho(1450) \\
 \omega(1420) \\
 \phi(1680) \\
 K^\ast(1680), K^\ast(1410)
 \end{array}\right) ~. 
\en
The candidates are rather well established, but the interpretation 
of the $K^\ast$ states is still an open issue: Experimentally, one finds 
$K^\ast(1680)$ and $K^\ast(1410)$ which would be possible candidates, 
even though the mass of $K^\ast(1410)$ (current PDG average is 
1414 $\pm$ 15 MeV) would be considerably low and possibly 
indicates mixing with a hybrid meson state in this mass region. 
In~\cite{Barnes:1996ff,Barnes:2002mu} the excited vector kaon state has been 
considered with mass 1414 MeV and 1580 MeV. We will also discuss
which decay patterns are expected for the $K^\ast=K(1680)$.

\subsubsection{Decays and fit}

We have performed fits to the $^3P_0$ data including both candidates for the 
excited vector kaon separately. The resulting coupling constants do not 
depend on this choice. 

We do, however, not include the very narrow decay width 
$\Gamma(K^\ast(1580) \rightarrow \eta' K)$ in our fit as this value seems 
to be a breakout. 
In Table \ref{table:totwidths}, we give the 
experimental total decay widths and the sum of the widths of the modes 
considered here, in the chiral approach and the results by 
Barnes {\it et al.}~\cite{Barnes:1996ff,Barnes:2002mu}. 
The results for the partial decay widths of the first vector radial excitation
are listed in Tables \ref{table:v1pp} and 
\ref{table:v1vp}. The results obtained in~\cite{Volkov:1997dd} lie 
considerably below both the $^3P_0$ model predictions and consequently 
our fit. Indeed, the experimental total decay width is overestimated by 
the two latter approaches. The candidates for the 
kaon are discussed below. One can expect that the decay modes not 
considered here (such as decays to scalar mesons) are important for the 
$\rho$ and weaker for the $\omega$ and $\phi$.

\begin{table}
\begin{center}
\caption{Total decay widths of first vector radial excitation (in MeV).} 
\label{table:totwidths}
\begin{tabular}{|c|c|c|c|} \hline
State & Data~\cite{Amsler:2008zz} 
& Sum of widths (Barnes {\it et al.}~\cite{Barnes:1996ff,Barnes:2002mu})  
& Sum of widths (Chiral approach) \\ 
\hline
$\rho(1450)  $& 400 $\pm$ 60 & 275 & 330 \\
$\omega(1420)$& 180 $-$ 250       & 376 & 454 \\
$\phi(1680) $&  150 $\pm$ 50 & 378 & 361 \\ \hline
\end{tabular}
\end{center}
\end{table}

\begin{table}
\begin{center}
\caption{Decay widths of first vector radial excitation to two pseudoscalars 
(in MeV).} 
\label{table:v1pp}

\begin{tabular}{|cc|c|c|c|c|} \hline
Decay mode &    
& Gerasimov {\it et al.}~\cite{Gerasimov:1981gj} 
& Volkov {\it et al.}~\cite{Volkov:1997dd} 
& Barnes {\it et al.}~\cite{Barnes:1996ff,Barnes:2002mu} 
& Chiral approach \\ 
\hline 
$\rho(1450)$	& $\rightarrow \pi \pi$   			
& 7 & 22 &  74  & 108   \\ 
	$ $	
& $\rightarrow K \overline K $  
&-&-&  35  & 23    \\ 
\hline
$	\omega(1420) $	& $\rightarrow K \overline K $   
&-&-& 31 & 19    \\ 
\hline
$ \phi(1680) $  & $\rightarrow K \overline K $   
&-& 10 & 89 & 91  \\ 
\hline 
$ K^\ast(1414)$  & $ \rightarrow \pi K $  			 
&-& 20 & 55 & 50    \\ 
& $ \rightarrow \eta K $  		 
&-& - & 42 & 23    \\ 
\hline
$ K^\ast(1580)$  & $ \rightarrow \pi K$   			 
&-&-& 61 & 76    \\ 
& $ \rightarrow \eta K$   		 
&-&-& 60 & 44    \\ 
& $ \rightarrow \eta^\prime K$  		 
&-&-& 0.5 & 6  \\ 
\hline
\end{tabular}
\end{center}
\end{table}
Note that in~\cite{Gerasimov:1981gj} the 
$\rho^\ast$ meson mass was placed at 1220 MeV. In this approach the decay 
width $\rho^\ast \rightarrow \pi \pi $ is strongly suppressed as a result 
of the node structure of the wave function. 

For the two pseudoscalar modes the coupling strength results are 
\eq
c_{_{V^\ast PP}} = 1.65 \ {\rm GeV}^{-1} \,. 
\en 

\begin{table} 
\begin{center}
\caption{Decay widths of first vector radial excitation to 
pseudoscalar and vector mesons (in MeV)}\label{table:v1vp}
\begin{tabular}{|cc|c|c|c|c|} \hline
Decay mode &     
&   Volkov {\it et al.}~\cite{Volkov:1997dd} 
& Barnes {\it et al.}~\cite{Barnes:1996ff,Barnes:2002mu} 
& Chiral approach \\  
\hline
$\rho(1450)$	& $\rightarrow \omega \pi $ 
& 75 & 122 & 165  \\ 
$$	& $\rightarrow \rho \eta  $   
&-& 25 & 19  \\ 
$$ & $\rightarrow K K^\ast $   
&-& 19 & 15   \\  
\hline
$ \omega(1420) $ & $\rightarrow K \overline {K^\ast} $   
&-& 5 & 4   \\ 
$ $ & $\rightarrow \rho \pi $   
& 225 & 328 & 422   \\ 
$ $ & $\rightarrow \omega \eta $   
&-& 12 & 9   \\ 
\hline	
$ \phi(1680) $ & $\rightarrow K K^\ast $
 & 90 & 245 & 241  \\ 
& $ \rightarrow \eta \phi $ & 	
-& 44 & 29     \\ 
\hline
$ K^\ast(1414)$  & $ \rightarrow \omega K $   		
&-& 10 & 8    \\ 
& $ \rightarrow \rho K   $   		
&-& 34 & 26    \\ 
& $ \rightarrow  \pi K^\ast $   		
&-& 55 & 63    \\ 
& $ \rightarrow  \eta K^\ast$   		
&-& 0 &  0  \\  \hline
$ K^\ast(1580)$  & $ \rightarrow \omega K $   		
&-& 29 & 29    \\ 
& $ \rightarrow \rho K   $   		
&-& 90 & 91    \\ 
& $ \rightarrow \pi K^\ast  $   		
&-& 99 & 135    \\ 
& $ \rightarrow \eta K^\ast $   		
&-& 1 & 28  \\ 
& $ \rightarrow  \phi K  $   		
&-& 9 & 6    \\  
\hline
\end{tabular}
\end{center}
\end{table}

Again, for the pseudoscalar--vector modes we do not fit the very small decay 
widths. The coupling strength results in this case are 
\eq 
c_{_{V^\ast VP}}= 1.20 \ {\rm GeV}^{-1} \,. 
\en 

\subsubsection{The kaon}

The actual experimental candidates for the vector kaon are the 
$K^\ast(1680)$ and the $K^\ast(1410)$ with masses $1717\pm 27$~MeV, 
$1414 \pm 15$~MeV and total widths of $322\pm 110$~MeV and $232\pm21$~MeV,  
respectively. The $K^\ast(1414)$ lies somewhat too low with respect to 
the other better established members of the nonet, $K^\ast(1680)$ 
(with a mass of 1717~MeV) has a very high mass, above the $\phi (1680)$.

Table \ref{table:kaon1416} 
shows the known decay widths of $K^\ast(1410)$ and $K^\ast(1680)$ 
and our predictions in the chiral approach. 
\begin{table} 
\begin{center} 
\caption{Decay widths of $K^\ast$ candidates 
and predictions (in MeV).}\label{table:kaon1416}
\begin{tabular}{|c|c|c||c|c|} \hline
Decay mode 
& $K^\ast(1410)$ Data~\cite{Amsler:2008zz}  
& $K^\ast(1410)$ Chiral approach 
& $K^\ast(1680)$ Data~\cite{Amsler:2008zz}  
& $K^\ast(1680)$ Chiral approach\\  
\hline 
$K\pi$ & 15 &	52 & 125 & 109  \\ 
\hline
$K \eta$  & - &	26 & - & 50 \\ 
\hline
$K \eta^\prime$ & - & $\approx$ 0  & - & 0.5 \\ 
\hline 
$K \omega$ & - & 14 & -	& 58 \\ 
\hline
$K\rho$	& $< $16 & 43 &	101 & 178 \\ 
\hline
$\pi K^\ast $ & 	$ > $93	& 76    
& 96 &	221 \\  
\hline
$  \eta K^\ast $ & - &		13    
& - & 103 \\ 
\hline
$ \eta' K^\ast $ & - & $\approx$ 0 & - & 0.4 \\ 
\hline
$ K \phi $ &	- & 4 &	- & 49	\\ 
\hline 
\end{tabular}
\end{center}
\end{table}
Neither candidate is described well by the chiral approach. 
Although the decay ratio 
\eq 
\frac{\Gamma(K^\ast(1680)\rightarrow K\rho) }{ \Gamma(K^\ast(1680) 
\rightarrow K^\ast \pi)} \approx 1 
\en  
is reproduced approximately by the model, the total width of 
$K^\ast(1680)$ is largely overestimated, as observed before for $\omega$ 
and $\rho$.

The deviation from the experimental data is still stronger for the case of 
the $K^\ast(1414)$, where neither the current pattern of the decay widths
can be reproduced nor the decay width 
to $K\pi$. The decay widths to two pseudoscalars was not overestimated 
so strongly to explain a discrepancy by a factor 4.

Further measurements of $K \eta$ and $K^\ast \eta$ would help to confirm 
or disprove the interpretation of $K^\ast(1680)$ as the first radially 
excited vector kaon. The interpretation of the kaon remains an open issue. 
In this vein we should stress that SU(3) flavor symmetry
breaking effects could be 
quite important for modes involving strange mesons. This can be accommodated 
in the chiral approach by including the terms in the Lagrangian responsible 
for such symmetry breaking contributions. We plan to make this improvement 
of the approach in further work, where the presence of sufficient
data allows such a detailed analysis.

\subsubsection{The $\phi$ and the $\omega$}

The dominance of the $\pi\rho$ channel for the $\omega$ and the 
$K^\ast K$ channel for the $\phi$ agree well with experiment:
these two decays are stated by the PDG as the dominant ones.
The $^3P_0$ model as well as our fit overestimate
largely the total widths, especially in the vector--pseudoscalar channel.
The dominance of the $K \overline{K^\ast}$ mode, which is stated 
by PDG for the experimental candidate, is confirmed in the model 
and the fit. The interpretation of $\phi(1680)$ as a dominant
$s\bar s$ partner of the $\omega(1420)$ is therefore clear.
The width of this meson is estimated by PDG to be $215 \pm 35$ MeV, 
the sum of widths in~\cite{Barnes:1996ff,Barnes:2002mu} and in our 
approach lie considerably above that.

A more quantitative measurement of the widths of the single decay 
modes would help to improve the quality  of the present interpretation 
considerably. For example our estimate for 
$\frac{\Gamma(\phi \rightarrow K \overline K)}
{\Gamma(\phi \rightarrow K \overline{K^\ast})}\approx 
\frac 1 3$ is far from being consistent with the actual 
(not confirmed) value of $0.07 \pm 0.01$.

\section{Summary}\label{sec:end}

We have studied the strong decays of radially excited mesons in a chiral 
approach. Because of the lack of sufficient data points we chose to adjust 
the phenomenological coupling strengths of the present chiral approach  
for the decays of radial pseudoscalar and vector excitations to the 
average values of the $^3P_0$ model of 
Refs.~\cite{Barnes:1996ff,Barnes:2002mu}. 
Although the absolute values of the decay widths are fixed in such a fit,
the relative decay rates are a prediction of the chiral approach
when staying both in the initial and final states within fixed meson nonets.
Our goal was to test the possibility of a phenomenological analysis within 
such an approach, which would have also the advantage of being able to 
incorporate three--body decays.

The picture we have drawn is satisfactory. Most of the presently known decay 
widths are 
reproduced rather well, especially when taking into account the small number 
of parameters. A chiral approach analysis might therefore help to resolve 
the remaining interpretation issues in the future. The possibility to include 
three--body decays in our analysis might help to better understand the 
decay pattern, for example, in the context of the $\eta(1405)$, 
$\eta(1475)$ puzzle. Open interpretation issues were addressed but could 
not be resolved unambiguously, since the lack of experimental data prohibits 
a direct fit of our parameters. In the pseudoscalar sector an interesting 
extension is also given by the possible presence of a glueball state, which 
can mix with the quarkonia configurations.
The phenomenological consequences of this additional glueball configuration
on the decay patterns of pseudoscalar mesons are presently~studied. 

\begin{acknowledgments}

This work was supported by the DFG under Contract Nos. FA67/31-1, 
No. FA67/31-2, and No. GRK683. This research is also part of the EU
Integrated Infrastructure Initiative Hadronphysics project under
Contract No. RII3-CT-2004-506078 and President of Russia
"Scientific Schools"  Grant No. 871.2008.2. 

\end{acknowledgments}

\appendix 
\section{Matrices $V_{\mu\nu}$, $V_{\mu\nu}^\ast$, 
$P^\ast$ and $P^{\ast\ast}$} 

\eq 
V_{\mu\nu} &=& 
\left(\begin{array}{ccc}
 \displaystyle\frac{\rho^0}{\sqrt{2}} 
+ \displaystyle\frac{\omega}{\sqrt{2}}  
& \rho^+ & K^{\ast \, +} \\ 
 \rho^-& - \displaystyle\frac{\rho_0}{\sqrt{2}} 
+ \displaystyle\frac{\omega}{\sqrt{2}} & K^{\ast 0}\\
 K^{\ast \, -} & \overline{K^{\ast \, 0}} & \phi  
 \end{array}\right)_{\mu\nu} \,, \\[3mm]  
V_{\mu\nu}^\ast &=& 
\left(\begin{array}{ccc}
 \displaystyle\frac{\rho^0(1450)}{\sqrt{2}} 
+ \displaystyle\frac{\omega(1420)}{\sqrt{2}} & \rho^+(1450)
& K^{\ast \, +}(1680) \\ 
 \rho^-(1450)& - \displaystyle\frac{\rho^0(1450)}{\sqrt{2}} 
+ \displaystyle\frac{\omega(1420)}{\sqrt{2}} & K^{\ast 0}(1680)\\
 K^{\ast \, -}(1680) & \overline{K^{\ast \, 0}}(1680) & \phi(1680)   
 \end{array}\right)_{\mu\nu} \, \\\
P^\ast &=& \left(\begin{array}{ccc}
  \displaystyle\frac{\pi^0(1300)}{\sqrt{2}} +\eta(1295) 
& \pi^+(1300) & K^+(1460)\\
 \pi^-(1300)& - \displaystyle\frac{\pi^0(1300)}{\sqrt{2}} 
 + \eta(1295) & K^0(1460)\\
 K^-(1460)& \overline{K^0}(1460) &  \eta(1475) 
 \end{array}\right)  \\
 P^{\ast\ast} &=& \left(\begin{array}{ccc}
  \displaystyle\frac{\pi^0(1800)}{\sqrt{2}} + \eta(1760) 
& \pi^+(1800) & K^+(1830)\\
  \pi^-(1800) & - \displaystyle\frac{\pi^0(1800)}{\sqrt{2}} 
+ \eta(1760)  & K^0(1830)\\
 K^-(1830)& \overline{K^0}(1830) & 
\eta(2225) 
\end{array}\right)
\en 

\newpage 

\section{Tree--level amplitudes for the decay of excited mesons} 
\label{app:tables}
The tree--level amplitudes can easily be read off the original 
Lagrangian. We state them here for future reference. 

(1) Decay amplitudes of excited pseudoscalar mesons are given in Table XI. 
 
\begin{table}[hbt]
\begin{center}
\caption{Decay amplitudes squared (excluding the decay strength constant) } 
\def\arraystretch{1.5}
\begin{tabular}{|c|c|c|}\hline
Decay & Products & Squared amplitude \\ 
\hline 
$\pi^\ast\rightarrow P \, V$ & $\pi \rho$  &  $4$    \\ 
& $ K \overline{ K^\ast} $ &    $   2 $      \\ 
\hline 
$\pi^\ast\rightarrow V \, V$        & $\rho \phi$	  	
&     $ 8 \cos^2 \phi_V$         \\ 
& $\rho \omega$	& $ 8 \sin^2 \phi_V $         \\ 
& $K^\ast \overline{K^\ast}$ &      4       \\ 
\hline 
$ {\eta^{\ast}}^\prime \rightarrow P \, V $ & $ K \overline {K^\ast} $ &  
$ 6 \sin^2 \theta_P$\\ 
\hline 
$ {\eta^{\ast}}^\prime \rightarrow V \, V $ & $\rho^0 \rho^0  $ 
& $  2 \sin^2 \phi_P $\\ 
 & $\rho^+ \rho^-  $&  $ 8\sin^2 \phi_P$\\ 
 & $\phi \phi  $ &   
$   2 \left( \sin \phi^\ast_P \cos \phi_V^2 
+ \sqrt 2 \cos \phi^\ast_P \sin \phi_V^2 \right)^2 $\\ 
& $\phi \omega  $ &  $  8 \sin^2 \phi_V \cos^2 \phi_V 
\left( -\sin \phi^\ast_P + \sqrt 2 \cos \phi^\ast_P \right)^2$\\ 
& $\omega \omega $  & $  2 \left( \sin \phi^\ast_P \sin \phi_V ^2 
+ \sqrt 2 \cos \phi^\ast_P \cos \phi_V^2 \right)^2$\\ 
 & $K^\ast \overline{K^\ast}$  &  $  2 \left( \sin \phi^\ast_P 
+ \sqrt 2 \cos \phi^\ast_P \right)^2$\\ 
\hline 
$K^\ast \rightarrow P \, V$ & $ K \rho $      &  $  \frac 3 2  $  \\  
& $ K \omega $ & $ {\frac 3 2} \sin^2 \theta_V $  \\  
& $ K \phi   $ & $ {\frac 3 2} \cos^2 \theta_V $  \\ 
& $ \eta K^\ast $     &  $ {\frac 3 2} \cos^2 \theta_P $  \\  
& $ \eta^\prime K^\ast $ & $ {\frac 3 2} \sin^2 \theta_P $  \\  
& $ \pi K^\ast $  &  $ \frac 3 2  $  \\
 \hline 
$ K^\ast \rightarrow V \, V $ & $ K^{\ast} \rho $&  $ 6 $  \\ 
& $ K^\ast \omega $ & $ 4 (\sin \theta_V + \sqrt 2 \cos \theta_V)^2$ \\ 
& $ K^\ast \phi   $ & $ 4 (\cos \theta_V -\sqrt 2 \sin \theta_V)^2$ \\ 
\hline 
\end{tabular}
\end{center}
\end{table} 
The amplitudes for $\eta^\ast$ can be obtained easily by exchanging 
$\sin \rightarrow \cos, \cos \rightarrow - \sin$.

\newpage 

(2) Decay amplitudes of excited vector mesons are given in Table XII. 

\begin{table}[hbt]  
\begin{center}
\caption{Decay amplitudes squared (excluding the decay strength constants)} 
\def\arraystretch{1.5}
\begin{tabular}{|c|c|c|}\hline
Decay & Products & Squared amplitude \\ 
\hline 
$ {\rho^\ast} \rightarrow P \, P $ & $ K \overline {K} $& 4 \\  
 & $ \pi \pi  $ &$ 8 $\\ \hline 
$ {\rho^\ast} \rightarrow V \, P $ & $ \overline{K^\ast} K  $& 2 \\ 
 & $ \rho \eta   $&$  2  \cos^2 \phi_P $\\ 
 & $ \rho \eta^\prime   $&$ 2 \sin^2 \phi_P $\\ 
 & $ \omega \pi   $&$ 2 \sin^2 \phi_V $\\ 
 & $ \phi \pi   $ &$ 2 \cos^2 \phi_V $\\  
\hline 
$ {\omega^\ast} \rightarrow P \, P $ & $ K\overline K  $ 
& $  12  \sin^2 \theta^\ast_V $ \\ 
\hline 
$ {\omega^\ast} \rightarrow V \, P $ & $ \phi \eta $&$  
2\left( \sqrt 2 \cos \phi^\ast_V \sin \phi_V \sin \phi_P 
+ \sin \phi^\ast_V \cos \phi_V \cos \phi_P \right)^2 $\\ 
& $ \phi \eta^\prime $& $ 2\left( \sqrt 2 \cos \phi^\ast_V 
\sin \phi_V \cos \phi_P - \sin \phi^\ast_V 
\cos \phi_V \sin \phi_P \right)^2 $\\  
 & $ \omega \eta $&$  2\left( \sqrt 2 \cos \phi^\ast_V 
\cos \phi_V \sin \phi_P - \sin \phi^\ast_V \sin \phi_V 
\cos \phi_P \right)^2 $\\  
& $ \omega \eta^\prime $ & $  2\left( \sqrt 2 \cos \phi^\ast_V 
\cos \phi_V \cos \phi_P + \sin \phi^\ast_V \sin \phi_V 
\sin \phi_P \right)^2$\\  
 & $ \rho \pi $&$ 6 \sin^2 \phi^\ast_V $\\ 
 & $ K^\ast \overline K  $&$ 2\left( \sin \phi^\ast_V + \sqrt 2 
\cos \phi^\ast_V \right)^2 $\\ \hline 
$ {\phi^\ast} \rightarrow P \, P $ & $ K\overline K  $ 
& $ 12 \cos^2 \theta^\ast_V $\\ \hline 
$ {\phi^\ast} \rightarrow V \, P $ & $ \phi \eta $ 
& $  2\left( -\sqrt 2 \sin \phi^\ast_V \sin \phi_V \sin \phi_P 
+ \cos \phi^\ast_V \cos \phi_V \cos \phi_P \right)^2 $\\  
 & $ \phi \eta^\prime $& $  2\left( \sqrt 2 \sin \phi^\ast_V 
\sin \phi_V \cos \phi_P + \cos \phi^\ast_V \cos \phi_V \sin 
\phi_P \right)^2 $\\ 
 & $ \omega \eta $& $   2\left( \sqrt 2 \sin \phi^\ast_V 
\cos \phi_V \sin \phi_P + \cos \phi^\ast_V \sin \phi_V 
\cos \phi_P \right)^2 $\\  
 & $ \omega \eta^\ast $& $  2\left( -\sqrt 2 \sin \phi^\ast_V 
\cos \phi_V \cos \phi_P + \cos \phi^\ast_V \sin \phi_V 
\sin \phi_P \right)^2$\\ 
 & $ \rho \pi $& $  6 \cos^2 \phi^\ast_V $\\ 
 & $ K^\ast \overline K  $& $ 2\left( \cos \phi^\ast_V 
- \sqrt 2 \sin \phi^\ast_V \right)^2 $\\ 
\hline 
$ K^\ast \rightarrow P \, P$ & $ K \pi $  & $ 6  $ \\ 
& $ K \eta $ &  $ 4 (\frac 1 {\sqrt 2} \cos \phi_P + \sin \phi_P)^2 $ \\  
 & $ K \eta^\prime $ &  $ 4 (\frac 1 {\sqrt  2} 
\sin \phi_P - \cos \phi_P)^2$ \\ 
\hline 
$ K^\ast \rightarrow V \, P $ & $ K^\ast \pi $ & $ {\frac 3 2}  $ \\ 
 & $ K^\ast \eta $ & $ ( \frac {\cos \phi_P}{\sqrt 2} 
+ \sin\phi_P )^2  $ \\ 
 & $ K^\ast \eta^\prime	$&$ ( \frac {\sin \phi_P}{\sqrt 2} 
- \cos\phi_P )^2  $ \\ 
& $ \omega K $ & $ ( \frac {\sin \phi_V}{\sqrt 2} + \cos\phi_V )^2  $ \\ 
& $ \phi K $ & $ ( \frac {\cos \phi_V}{\sqrt 2} - \sin\phi_V )^2  $ \\ 
& $ \rho K $ &  $ 3  $ \\ 
\hline
\end{tabular}
\end{center}
\end{table}

\end{document}